\documentclass[twocolumn,showpacs,preprintnumbers,amsmath,amssymb]{revtex4}
\usepackage{graphicx}
\usepackage{dcolumn}
\usepackage{bm}
\usepackage{amsfonts}

\begin{document}

\preprint{APS/123-QED}

\title{Classical phase-space approach for coherent matter waves}

\author{Fran\c{c}ois Impens$^{1,2}$, David Gu\'ery-Odelin$^{3}$}

\affiliation{$^{1}$ SYRTE, Observatoire de Paris, CNRS, 61 avenue de
l'Observatoire, 75014 Paris, France}

\affiliation{$^{2}$ Instituto de Fisica, Universidade Federal do Rio
de Janeiro. Caixa Postal 68528, 21941-972 Rio de Janeiro, RJ,
Brasil}

\affiliation{$^{3}$ Universit\'e de Toulouse ; UPS ; Laboratoire
collisions Agr\'egats R\'eactivit\'e, CNRS ;  F-31062 Toulouse,
France}

\date{\today}

\begin{abstract}
We investigate a classical phase-space approach of matter-wave
propagation based on the Truncated Wigner Equation (TWE). We show
that such description is suitable for ideal matter waves in
quadratic time-dependent confinement as well as for harmonically trapped Bose Einstein condensates in the Thomas-Fermi regime. In
arbitrary interacting regimes, the TWE combined with the moment method
yields the low-energy spectrum of a condensate as predicted by
independent variational methods. TWE also gives the right
breathing mode frequency for long-ranged interactions
decaying as $1/r^2$ in 3D and for a contact potential
in 2D. Quantum signatures, beyond the TWE, may only be found in the condensate dynamics beyond the regimes of classical phase-space propagation identified here.
\end{abstract}


\maketitle

Introduced in the early days of quantum mechanics, Wigner distributions have been applied
successfully since then in many areas of physics. In optics it is used to describe partially coherent light
beams~\cite{Wigneroptics}, and in atomic
physics to characterize atom interferometers \cite{WignerInterfero} or Bose-Einstein condensates(BECs)~\cite{Steel98}.
The Wigner distribution obeys a propagation equation involving a series of
differential operators weighted by increasing powers of
$\hbar$~\cite{Walls94} which can be identified as the quantum terms of the Wigner equation. The classical limit of this equation, obtained by taking the limit $\hbar \rightarrow 0$, is referred to
as the Truncated Wigner Equation (TWE). This equation has been used with stochastic classical fields to study quantum gases~\cite{StochasticWigner}. We propose here to use it instead with the modes of a coherent atomic beam. When
combined with the moment method, this reveals an accurate yet simple way to capture the low-energy dynamics of a BEC. It is for instance remarkable that the TWE describes correctly the time-of-flight
expansion of condensates in the limit of vanishing interactions as well as in the Thomas-Fermi limit. Our treatment, which studies the agreement between the TWE and the full Wigner equation, also identifies the circumstances under which
quantum signatures -i.e. effects due to the quantum terms of the Wigner equation - may appear in the dynamics of BECs.

 After a quick reminder on the general Wigner equation,
we study its applications to the propagation of ideal matter waves in time-dependent quadratic
potentials. We retrieve the $ABCD$ propagation formalism~\cite{BordeABCD} for coherent guided atom optics~\cite{CoherentGuidedAtomOptics}, suitable to investigate
 atomic beams propagating in time domain~\cite{ASD96} through a variety of atom-optical elements. It can also be used
to study the transverse stability~\cite{Impens09ABCD} of an atomic resonator~\cite{matterwaveresonators}, to define the quality factor of an atomic beam~\cite{QualityFactor}, or to investigate the generation of atom-optical caustics~\cite{QualityFactor} caused by
a sudden potential change. We also analyse the predictions of the
TWE for the expansion and for the low-energy spectrum of interacting atomic waves evolving in harmonic traps. The
consistency of the TWE with a universal prediction for long-ranged
$1/r^2$ interactions is verified.\\

 We recall the form of the Wigner Equation in the
presence of a general two-body potential $V( \mathbf{r})$. The
non-relativistic atomic field operator $\hat{\psi}$ obeys the
general propagation equation
\begin{eqnarray}
\label{eq:equation champ interaction portee finie une dimension} i
\hbar \frac {\partial \hat{\psi}} {\partial t} ( \mathbf{r} ,t ) &
= & - \frac {\hbar^2 } {2 m} \Delta \hat{\psi} ( \mathbf{r} ,t )+
U_0 ( \mathbf{r},t)
\hat{\psi} ( \mathbf{r} ,t ) \nonumber \\
& + & \int {\rm d}  \mathbf{r}' \hat{\psi}^{\dagger}(
\mathbf{r}',t) V( \mathbf{r}-\mathbf{r}') \hat{\psi}(
\mathbf{r}',t) \hat{\psi}( \mathbf{r},t) \,,
\end{eqnarray}
where $U_0(\mathbf{r} ,t)$  is the time-dependent external
potential experienced by the atoms. We assume the sample to be
well-described by the mean-field approximation and in a number
state associated with a macroscopic single mode
$\phi$. The associated Wigner distribution is
\begin{equation}
 W(\mathbf{r},\mathbf{p},t)=  \int \frac {{\rm d} \mathbf{r}'} {(2 \pi \hbar)^3} \phi \left(\mathbf{r}+\frac {\mathbf{r}'} {2},t\right) \phi^{*}\left(\mathbf{r}-\frac {\mathbf{r}'}
 {2},t\right) e^{- \frac i \hbar  \mathbf{p} \cdot \mathbf{r}'} \, .
\end{equation}
The mode $\phi$ satisfies a Schr\"odinger equation with an
effective Hamiltonian $H(\mathbf{r},\mathbf{p},t)= \mathbf{p}^2 / 2 m+ U( \mathbf{r},t)$.
where  $U( \mathbf{r},t)= U_0( \mathbf{r},t)+U^{\rm{mf}}( \mathbf{r},t)$ and $U^{\rm{mf}}( \mathbf{r},t)=\int {\rm d} \mathbf{r}' |\phi(\mathbf{r}',t)|^2
V( \mathbf{r}-\mathbf{r}')$. 
For a contact interaction potential $V(\mathbf{r})=g \delta (\mathbf{r})$,
this Schr\"odinger equation reduces to the usual time-dependent
Gross-Pitaevskii equation(GPE). The Wigner distribution associated
with a wave-function governed by an Hamiltonian
$H(\mathbf{r},\mathbf{p},t)$ satisfies
 the transport equation \cite{Groot72,Wigner84} $\partial W / \partial t = - i {\cal L}
 \left[ W \right]$, where $\cal L$ is the Liouvillian operator
\begin{eqnarray}
\label{eq:Liouvillien quantique une dimension}{\cal L} & = &
H(\mathbf{r},\mathbf{p},t) \left[  \frac {2i} {\hbar} \sin \frac
{\hbar} {2} \overleftrightarrow{\Lambda} \right]
\nonumber \\
\quad & \mbox{with} & \quad \overleftrightarrow{\Lambda}=
\sum_{\eta=1,2,3} \frac {\overleftarrow{\partial}} {\partial
r_{\eta}} \frac {\overrightarrow{\partial}} {\partial p_{\eta}}-
\frac {\overleftarrow{\partial}} {\partial p_{\eta}} \frac
{\overrightarrow{\partial}} {\partial r_{\eta}} \, .
\end{eqnarray}
We have identified the vectors $\mathbf{r}=(x,y,z)$ and
$\mathbf{p}=(p_x,p_y,p_z)$ with $(r_1,r_2,r_3)$ and
$(p_1,p_2,p_3)$. The differentiation operators
$\overrightarrow{\partial}$ and $\overleftarrow{\partial}$
act on the Wigner distribution
$W(\mathbf{r},\mathbf{p},t)$ and on the Hamiltonian
$H(\mathbf{r},\mathbf{p},t)$ respectively. The Wigner equation can be recast in an explicit differential series
ordered by increasing powers of the constant $\hbar$
\begin{eqnarray}
\label{eq:Full Wigner Equation} & \: &  W_t
=  \sum_{i=1}^{3} \left( - \frac {p_{i}} {m} \: W_{r_{i}}+
U_{r_{i}} W_{p_{i}} \right) \nonumber \\  & +  &
\small \sum_{\substack{\small {n=1},
\\{n_1+n_2+n_3}={2n+1} \normalsize}}^{+\infty} \frac {(i \hbar)^{2n}} {2^{2n}
n_1!n_2!n_3!}    U_{r_1^{n_1} r_2^{n_2} r_{3}^{n_3}} W_{p_1^{n_1}
p_2^{n_2} p_3^{n_3}} \normalsize \,.
\end{eqnarray}
To alleviate notations, from now on we note the partial differentiation with respect to time, space and
momentum coordinates thanks to an index (i.e. $W_{p_1^{n_1} p_2^{n_2}
p_3^{n_3}}$ is the derivative of the Wigner distribution $\frac
{\partial^{n_1}} {\partial p_1^{n_1}} \frac {\partial^{n_2}}
{\partial p_2^{n_2}} \frac {\partial^{n_3}} {\partial p_3^{n_3}} W
$). The TWE is obtained by taking the classical limit $\hbar
\rightarrow 0$, i.e. by discarding the last term of the
right-hand-side (r.h.s.) in Eq.\eqref{eq:Full Wigner Equation}.

Let us first treat the propagation of ideal atomic waves in time-dependent quadratic potentials. In this regime, the
full Wigner equation exhibits no quantum terms and thus reduces to the TWE. We show that the Wigner equation, or here equivalently the TWE,
can be used to recover exactly the $ABCD$ propagation formalism. This phase-space method had been already verified by direct integration for 1D Laguerre-Gaussian modes \cite{BordeHouches}, and analogously for partially coherent
matter waves \cite{chenABCDarXiv07}. The Wigner equation offers a compact proof of this result, avoiding the computation of propagation integrals. 

Consider a general quadratic Hamiltonian (without atomic
interactions) expressed with the conventions of~\cite{BordeABCD}
\begin{equation}
\label{eq:Hamiltonien general}
H = \frac{ {^T\!\mathbf{p}} \cdot \beta \cdot \mathbf{p}} {2 m} -
{^T\!\mathbf{r}} \alpha \mathbf{p} - \frac {m} {2}
{^T\!\mathbf{r}} \gamma \mathbf{r} - m \overrightarrow{g} \cdot
\mathbf{r}
 + \overrightarrow{f} \cdot \mathbf{p} \,,
\end{equation}
 $\alpha$, $\beta$ and $\gamma$ are $3\times 3$
time-dependent real matrices satisfying $^T\!\alpha=-\alpha$,$^T\!\beta=\beta$, $^T\!\gamma=\gamma$, and $\overrightarrow{f}$,$\overrightarrow{g}$ are time-dependent vectors. $^T$ stands for the matrix or for the vector transposition.
 One
sees readily that the full Wigner equation is here simply given by
its classical part (TWE) $y_{t} = -
H_{\mathbf{p}}(\mathbf{r},\mathbf{p}) \cdot y_{\mathbf{r}}+
H_{\mathbf{r}}(\mathbf{r},\mathbf{p}) \cdot y_{\mathbf{p}}$.  We introduce the phase-space vector
$\mathbf{R}(t) = ( \mathbf{r},
 \mathbf{p}/m  )$. We expect the
following phase space map~\cite{BordeABCD}
\begin{equation}
\label{eq:phase space map}
\mathbf{R}(t)=
\: M(t,t_0) \: \mathbf{R}(t_0)  + \mathbf{R}_0(t,t_0)
\end{equation}
$\mathbf{R}_0(t,t_0)= (   \overrightarrow{\xi},
\overrightarrow{\phi}  )$ is a source term to be determined later.
 $M(t,t_0)$ is a $6 \times 6$ matrix, referred to as the $ABCD$
matrix associated with the evolution between the initial time
$t_0$ and final time $t$:  $M(t,t_0)= [A|B,C|D] \:$. We used the matrix notation $M=[m_{11}|m_{12},m_{21}|m_{22}]$ -$m_{ij}$ being
the element of the $i_{\mbox{th}}$ row and $j_{\mbox{th}}$ column- and $A, B, C,D$ are $3\times3$ matrices depending on the pair of instants $(t,t_0)$. The $ABCD$ matrix satisfies the symplectic relation $M^{-1}= [{^T\!D}| -{^T\!B},{^T\!C}| {^T\!A}]$. The phase-space
map~\eqref{eq:phase space map} and the symplectic relation lead us to consider the following ansatz as a possible
solution to the TWE
\begin{eqnarray}
\label{eq:Wigner ABCD evolution} F(\mathbf{r},\mathbf{p},t) & = &
W
  \left(  \: {^T\! D} (\mathbf{r} - \overrightarrow{\xi}) \right. -
  \frac{1}{m}
{^T\! B} (\mathbf{p} - m \overrightarrow{\phi}) \: ,  \nonumber \\
 & & - \left. m {^T\! C} (\mathbf{r} - \overrightarrow{\xi}) +
{^T \!A} (\mathbf{p} - m \overrightarrow{\phi}) \:  , \: t_0
\right) \,.
\end{eqnarray}
 This ansatz indeed satisfies the Wigner
equation if the following relation is fulfilled for any
phase-space point 
 \begin{eqnarray}
 & \: &
T_1 \mathbf{r} \cdot W_{\mathbf{r}}^{0} +  T_2 \frac {\mathbf{p}}  {m} \cdot
 W_{\mathbf{r}}^{0} + m T_3   \mathbf{r} \cdot W_{\mathbf{p}}^{0}
 \nonumber \\
 & + & T_4 \mathbf{p} \cdot
 W_{\mathbf{p}}^{0}
  +   \overrightarrow{T}_5 \cdot
W_{\mathbf{r}}^{0}
 + m \overrightarrow{T}_6 \cdot W_{\mathbf{p}}^{0}
 =  0 \, . \nonumber \\
\label{eq:regroupement matrices ABCD} \nonumber
\end{eqnarray}
with
\begin{eqnarray}
T_1 & = & {^T \! \dot{D}} - {^T\! D} {^T \!\alpha} - {^T \!B} \gamma \,, \quad  T_2 = - {^T \! \dot{B}}
+ {^T \!D} {^T \!\beta} -
 {^T\! B} \alpha \nonumber \\
T_3 & = &-  {^T \! \dot{C}}  +  {^T \! C} {^T \! \alpha} + {^T
\! A} \gamma \,, \quad  T_4 =  {^T \! \dot{A}} - {^T \! C} \beta  - {^T \! A} {^T \! \alpha}  \nonumber \\
\overrightarrow{T}_5 & = & {^T \! \dot{B}} \overrightarrow{\phi}+  {^T \! B} \dot{\overrightarrow{\phi}}- {^T \! B} \overrightarrow{g} -({^T \! \dot{D}} \overrightarrow{\xi} - {^T \! D}
\dot{\overrightarrow{\xi}}+ {^T \! D} \overrightarrow{f}) \nonumber \\
\overrightarrow{T}_6 & = &   {^T \! \dot{C}} \overrightarrow{\xi}+ {^T \! C}
\dot{\overrightarrow{\xi}}- {^T \! C} \overrightarrow{f} -  ({^T \! \dot{A}} \overrightarrow{\phi} +^T \! A \dot{\overrightarrow{\phi}}- {^T \! A}
\overrightarrow{g})
\label{eq:regroupement matrices ABCD 2} \nonumber
\end{eqnarray}
The upper dots stand for time derivatives and lower dots denote scalar products. The terms $W_{\mathbf{r}}^{0},\: W_{\mathbf{p}}^{0}$
correspond to the gradients of the Wigner distributions towards
the position and momentum vectors respectively, evaluated at the
initial time $t_0$. It is now convenient to introduce the matrix $N$ and vector $S$ defined from the Hamiltonian~\eqref{eq:Hamiltonien general}
as  $N=[\alpha|\beta,\gamma|\alpha]$ and $\mathbf{S}=(\overrightarrow{f},
 \overrightarrow{g})$. Eq.\eqref{eq:regroupement matrices ABCD} implies
 that $T_{1,2,3,4}=0_3$ and $\overrightarrow{T}_{5,6}=\overrightarrow{0}$, i.e. the parameters $A,B,C,D,\overrightarrow{\xi}$, and $\overrightarrow{\phi}$
 satisfy the differential equations
\begin{equation}
\label{eq:systeme differentiel verifie ABCD}   \frac {{\rm d} M} {{\rm d} t} \:
 = N  M \,, \quad   \frac {{\rm d}} {{\rm d} t} \left( M^{-1} R_0
\right) = M^{-1} S \, ,
\end{equation}
with the initial conditions
 $\overrightarrow{\xi}(t_0,t_0)=\overrightarrow{\phi}(t_0,t_0)=\overrightarrow{0}$,
$A(t_0,t_0)=D(t_0,t_0)=I_{3}$ and $B(t_0,t_0)=C(t_0,t_0)=0_3$.
$I_{3}$ and $0_3$ denote respectively the identity and the zero
$3\times 3$ matrix. The system obtained in~\cite{BordeABCD} for
the parameters
$A,B,C,D,\overrightarrow{\xi},\overrightarrow{\phi}$ is equivalent
to (\ref{eq:systeme differentiel verifie ABCD}). With the adequate
initial values for
$A,B,C,D,\overrightarrow{\xi},\overrightarrow{\phi}$, the ansatz
$F$ coincides initially with the Wigner
 distribution and satisfies the same differential equation: the Wigner function thus evolves according to Eqs.~\eqref{eq:Wigner ABCD
 evolution} and~\eqref{eq:systeme differentiel verifie ABCD}.


The TWE also gives the right predictions for the expansion of a dilute BEC trapped in the
Thomas-Fermi limit, in particular the correct expansion law~\cite{CastinDum96} expected in a time-of-flight experiment and the corresponding low-lying mode frequencies~\cite{VarennaKetterle}. Here, we summarize only the main points of the argument. The BEC, subject to contact interactions, evolves in a time-dependent quadratic external potential $U(\mathbf{r},t)= \sum_{i=1}^{3} m \omega_i^2(t) r_i^2/2$.
Our approach combines the technique of scaling
factors already exploited by several
authors~\cite{CastinDum96,Kagan97,Michinel96} with the TWE. The transposition of scaling laws to Wigner functions consists in
searching for an ansatz of the form
$W(\mathbf{r},\mathbf{p},t)\simeq
\widetilde{W}(\mathbf{r}',\mathbf{p}',0)$
with a gauge transform~\cite{David02} $(r'_i,p'_i)= ( r_i/
b_i(t) , b_i(t) p_i - m \dot{b}_i(t) r_i)$. This ansatz depends on time only through the scaling parameters
$b_i$. Its adequacy with the exact Wigner distribution in
the Thomas-Fermi limit is not obvious and has not been considered so far
to our knowledge. For this purpose, one writes the Wigner distribution as
$W(\mathbf{r},\mathbf{p},t)=
\widetilde{W}(\mathbf{r}',\mathbf{p}',0)+\delta \widetilde{W}$,
and shows that the time-dependent correction $\delta
\widetilde{W}$ vanishes in this regime. Inserting this expression in the TWE, one sees that this correction is
driven by a term which vanishes if the scaling parameters satisfy
$\ddot{b}_i = \omega_i^2(0) /(b_i  \prod_{j=1}^3 b_j ) -
b_i \omega_i^2(t)$. Using a procedure suggested in Ref.~\cite{CastinDum96}, one can show that $\delta
\widetilde{W}$ remains negligible in the Thomas-Fermi limit.

\label{sec:moments}

A powerful theoretical technique to track the phase-space motion
is provided by the use of moments, and has been extensively used
in combination with non-linear Schr\"odinger-like equations, in
optics and electromagnetism~\cite{Talanov70,Porras93}. It turns
out to be useful also to derive low-lying collective modes of a
harmonically trapped Bose gas degenerate or not. In the following, we detail the transposition of this technique
within the Wigner's formalism.  Moments are defined with the
Wigner distribution by $\langle a(\mathbf{r},\mathbf{p}) \rangle(t)= \int {\rm
d}^3\mathbf{r} {\rm d}^3\mathbf{p} \: a(\mathbf{r},\mathbf{p}) \:
W(\mathbf{r}, \mathbf{p},t)$ with $a(\mathbf{r},\mathbf{p})$ a generic monomial in the position
and momentum coordinates. By convention we define the order of the
moment as the total power of the monomial. The orders in position
and momentum are the partial powers associated respectively with
the product of the position and the momentum coordinates inside
the moment.


It is insightful to investigate whether the predictions of the TWE regarding the dynamics of moments differ
from those of the full Wigner equation.  From the full
Wigner equation~\eqref{eq:Full Wigner Equation}, we deduce that the first-order time derivative of moments with an order
in momentum strictly less than three can be expressed exactly by
considering only the classical terms of this equation, that is the TWE. Indeed, considering the moment $\langle r_1^{a_1}
r_2^{a_2} r_3^{a_3} p_1^{b_1} p_2^{b_2} p_3^{b_3} \rangle$, one sees that its time derivative involves only the terms of the quantum series~\eqref{eq:Full Wigner Equation}
satisfying $n_k \leq b_k$ for
$k=1,2,3$. In particular, the first-order time derivative of any second-order moment is given exactly by the TWE.

For ideal matter waves propagating in quadratic potentials, the two terms in the r.h.s. of the TWE, namely $- p_{i}/m \: W_{r_{i}}$ and
$U_{r_{i}} W_{p_{i}}$, couple the motion of a moment only to other moments of equal or lesser order. With the remark above on the time derivative of moments, one obtains that the TWE describes exactly the dynamics of second-order moments.
Besides, any moment follows also a closed set of equations of motion. In contrast, if the effective potential $U$ contains polynomials beyond the second order - coming either from the external potential, either from mean-field interactions -, these properties are no longer verified. Indeed, the second classical term in the r.h.s. - $U_{r_{i}} W_{p_{i}}$ - couples the motion of any moment with a non-zero momentum order to higher-order moments. Second-order moments are then coupled to a hierarchy of moments of arbitrary high order, some of which influenced by the quantum terms of the Wigner equation. In this case, one must use the full Wigner equation to capture the moment dynamics.

By applying the TWE to specific moments which are not directly influenced by the quantum terms, it is still possible to infer exactly the low-energy mode frequencies of an interacting sample in the presence of either contact, dipolar or $1/r^2$ interactions. We focus from now on the motion of second-order moments. Their first-order time derivative can be evaluated exactly without considering the quantum terms in the Wigner equation. In this sense, the equations of motion~\cite{DavidRefMoments}
\begin{equation}
\label{eq:dynamics second order moments}
 \frac {\rm{d} \langle r_i^2 \rangle}  {{\rm d} t}  =  \frac {2} {m} \langle r_i  p_i \rangle  \,, \: \: \:
\frac {\rm{d} \langle r_i  p_i \rangle }  {{\rm d} t}  =   \frac
{1} {m} \langle p_i^2 \rangle - m \omega_i^2 \langle r_i^2 \rangle
- \langle r_i  U^{\rm{mf}}_{r_i} \rangle
\end{equation}
are classical and can be regarded as TWE predictions.

Let us prove that the TWE respects the universal ground frequency
invariance~\cite{Pitaevskii97} expected when one adds an
interaction potential $V$ satisfying the scaling law $V(\lambda \mathbf{r})=\lambda^{-2} V(\mathbf{r})$.
In a 3D system this property is only fulfilled by the long-ranged
$1/r^2$ potential, but in a 2D system (such as a condensate with a
frozen external degree of freedom) it is also valid for contact
potentials such as $g \delta(\mathbf{r})$. Similarly to the derivation of the virial theorem in mechanics, 
one obtains by differentiation of Eqs.\eqref{eq:dynamics second order moments}
\begin{equation}
\label{eq:invariantunsurr2 equation moment r2 2} \frac {m} {2}
\frac {{\rm d}^2 \langle r^2 \rangle } {{\rm d}t^2}=   \frac
{\langle p^2 \rangle} {m} - m \omega_0^2 \langle r^2 \rangle + \:
\langle \mathbf{r} \cdot U^{\rm{mf}}_{\mathbf{r}} \rangle \, .
\end{equation}
The scaling law of the
interaction potential implies that $\langle \mathbf{r} \cdot
U^{\rm{mf}}_{\mathbf{r}} \rangle  =  -2 \langle
U^{\rm{mf}} \rangle $. One obtains the closed equation
${\rm d}^2 \langle r^2 \rangle/{\rm d}t^2 + 4 \omega_0^2 \langle r^2
\rangle = 4 E/ m$ with the total energy $E= \langle p^2 \rangle /2 m+
 m \omega_0^2 \langle r^2 \rangle /2 + \langle
U^{\rm{mf}}\rangle$.
The lowest mode frequency $2 \omega_0$ is thus immune
to the addition of an interacting potential with the scaling property above.


 We now show that the classical Eqs.\eqref{eq:dynamics second order moments} are sufficient to predict the low-energy modes of an interacting condensate. We choose an interaction potential in the general
form~\cite{DipolarInteractions} $V(\mathbf{r})= g_c \delta( \mathbf{r})+g_d  (1 - 3 \cos^2
\theta_{r})/ r^2$. It contains a contact term and a dipolar contribution associated
with a permanent moment, dipoles are aligned along the vector $\mathbf{e}_3$ and $\cos \theta_{r}= (\mathbf{r} \cdot \mathbf{e}_3)/r$.
As such the TWE cannot be solved analytically, but we can investigate its predictions thanks to
a Gaussian ansatz
\begin{equation}
\label{eq:ansatz wigner} W(\mathbf{r},\mathbf{p},t)= \frac  {1}
{(\pi \hbar)^3} \prod_{i=1,2,3} e^{- \frac {(r_i-r_{0i})^2}
{\lambda_i^2}} e^{- \frac {\lambda_i^2} {\hbar^2} ( p_i  - \hbar
\alpha_i - 2 \hbar \beta_i \: r_i)^2}  
\end{equation}
$r_{0i}$ is the average position, $\lambda_i$ the width of the
wave-packet, and the parameters
$\alpha_{i},\beta_i$ define a quadratic wave-front. Using the relation ${\rm d} \langle r_i \rangle/ {\rm d}t = \langle p_i
\rangle / m$ and the first Eq.~\eqref{eq:dynamics second
order moments}, one finds $\alpha_i  =  (m/ \hbar) [ \dot{r}_{0i} - (\dot{\lambda}_i/ \lambda_i) r_{0i} ]$ and $\beta_i  = m  \dot{\lambda}_{i} /(2 \hbar \lambda_{i})$.
Using these expressions in Eq.\eqref{eq:ansatz wigner} yields the scaling gauge transform introduced previously. Eqs.\eqref{eq:dynamics second order moments} give for each coordinate an equation of motion similar to Eq.\eqref{eq:invariantunsurr2 equation moment r2 2}. The second-order moments in position and in momentum can be written with the ansatz parameters as
$\langle r_i^2 \rangle=  r_{0i}^2+ \lambda_i^2 /2$ and $\langle p_i^2 \rangle =  m^2  (  \dot{r}_{0i}^2
 +  \dot{\lambda}_i^2 /2 )  + \hbar^2 /(2 \lambda^2_i)$.
Reporting these expressions in the equations of motion, and using $\ddot{r}_{0i}=- \omega_i^2 r_{0i}$, one finds that
the Gaussian widths $\overrightarrow{\lambda}$ satisfy
\begin{equation}
\label{eq:equation wi 1} m   \ddot{\lambda}_i  = - m \omega^2_i \:
\lambda_i + \frac {\hbar^2} { m \lambda^3_i} - \: \frac {2}
{\lambda_i} \: \langle r_i \: U_{r_i}^{\rm{mf}} \rangle \,.
\end{equation}

 The contact potential contributes to the r.h.s. with $
\langle r_i U_{r_i}^{c} \rangle =  - g_c N / (4 \sqrt{2}
\pi^{3/2} \lambda_1 \lambda_2 \lambda_3)$, yielding a gradient in Eq.\eqref{eq:equation wi 1}. By setting $g_d=0$ and $\lambda_i=R_i^G b_i$ with adequate constants $R_i^G$, Eq.\eqref{eq:equation wi 1} coincide with the previous differential system for the parameters $b_i$ up to a term vanishing in the Thomas-Fermi limit. The Gaussian ansatz, successful in both the dilute and strongly interacting limits, is thus a valid interpolation between these regimes.  This also shows the resilience of surface modes towards the sample profile. The dipolar potential also contributes through a gradient,
thereby turning Eq.\eqref{eq:equation wi 1} into the equation of motion of a fictive point-like particle of mass $m$, of
position $\overrightarrow{\lambda}$, and experiencing an effective potential $V^e$
\begin{eqnarray}
\label{eq:wigner gaussian potentiel effectif} & \: &
V^e(\overrightarrow{\lambda}) = \sum_{i=1}^3 \left( \frac {\hbar^2}
{2 m \lambda^2_i} +\frac {m} {2} \omega^2_{i} \lambda_i^2 \right)  +
\frac { N } {\pi^{3/2} \lambda_1 \lambda_2 \lambda_3} \nonumber \\ &
\times & \left[
 \frac {g_c} {2 \sqrt{2}}
 +  4 \sqrt{2} g_d \int {\rm d} \mathbf{r}
\frac {1 - 3 \cos^2 \theta_{\mathbf{r}}} {r^2}  \prod_{j=1}^3 e^{-
r^2_i / (2 \lambda_i^2)} \right] \, . \nonumber \\
\end{eqnarray}
This result matches perfectly the predictions of the variational
methods~\cite{YiYou02,Michinel96}, in which the same differential
system $m \: \ddot{\lambda_i} \: = \: - \: {\rm d}
V^e(\overrightarrow{\lambda})/{\rm d} \lambda_i$ was obtained. One retrieves
the low-energy monopolar and quadrupolar frequencies for a cylindrical condensate obtained by other theoretical methods~\cite{Michinel96,Stringari96a} and confronted successfully to the experiments~\cite{VarennaKetterle}. The
method of moments combined with the TWE is thus entirely
equivalent to the variational method on this class of Gaussian
ansatz, but it nicely avoids the usual algebraic operations required in the latter to uncouple the Euler-Lagrange equations.\\

In conclusion, the classical phase-space evolution given
by the TWE is sufficient to explain many important features in
the dynamics of zero-temperature condensates. Using an adequate transformation of the Wigner distribution,
we have recovered the dynamics of an expanding BEC in the Thomas-Fermi regime. With the method of moments, we have obtained the correct low-energy spectrum of a condensate in the presence of contact and dipolar
interactions of arbitrary strength. Besides, the universal oscillation frequency in the
presence of a $1/r^2$ potential is also predicted by the
TWE: this suggests that the hidden symmetry of the full
equation~\cite{Pitaevskii97} is preserved despite the truncation of the Wigner equation. This
phase-space approach has allowed us to revisit the $ABCD$ law for the
propagation of dilute and partially coherent matter-waves, which
can be extended to account for mean-field interactions~\cite{Impens09ABCD}. Last, we have shown that the quantum terms of the Wigner equation do not affect the low-energy modes of trapped interacting condensates. Since quantum signatures appear in regimes where the TWE fails, the presented results suggest that neither the Thomas-Fermi nor the diluted limit are appropriate to evidence such signatures: these should be rather tracked with intermediate mean-field interactions or in high-order modes. Recently, it has been brought to our knowledge that the variational and moment methods have been also investigated in~\cite{PerezSIAM07}.

The authors thank Yvan Castin for helpful suggestions and Fabricio Toscano for comments. F.I. thanks Luiz Davidovich and Nicim Zagury for hospitality. This work was supported by DGA and CNRS.

\end{document}